\documentclass[journal ]{new-aiaa}
\usepackage[utf8]{inputenc}
\usepackage{textcomp}

\usepackage{amsmath,amssymb,graphicx,fullpage,color,subfigure,wrapfig,epstopdf,diagbox,mathtools}
\usepackage[titlenumbered,ruled]{algorithm2e}
\usepackage[figuresright]{rotating}

\usepackage{amsmath}
\usepackage[version=4]{mhchem}
\usepackage{siunitx}
\usepackage{longtable,tabularx}
\usepackage{trace}

\newcommand{\vo}[1]{\boldsymbol{#1}}
\newcommand{\omegab}{{\vo{\omega}}}

\title{Computationally Efficient Attitude Estimation with Extended $\mathcal{H}_2$ Filtering}

\author{Sunsoo Kim\footnote{Graduate Student and AIAA Student Member}}
\affil{Texas A\&M University, Electrical \& Computer Engineering, College Station, TX , 77840}
\author{Vaishnav Tadiparthi \footnote{Graduate Student and AIAA Student Member}}
\author{Raktim Bhattacharya \footnote{Associate Professor, and AIAA Associate Fellow}}
\affil{Texas A\&M University, Aerospace Engineering, College Station, TX , 77840}

\begin{document}

\maketitle

\section{Introduction}
Accurate state estimation using low-cost MEMS (Micro Electro-Mechanical Systems) sensors present on Commercial-off-the-shelf (COTS) drones has been reported to be a challenging problem \cite{eure2013application, gebre2004design, kada2016uav, weibel2015small}.
Most UAV systems use a combination of a gyroscope, an accelerometer, and a magnetometer to obtain measurements and estimate attitude.
Since measurements from these sensors are subjected to noise, bias, as well as magnetic disturbances,

a filtering framework that incorporates sensor fusion is essential to obtaining a reliably accurate estimate of the vehicle's attitude  \cite{crassidis2007survey,ludwig2018comparison,teague2016comparison}.

Under the sensor fusion paradigm, the Extended Kalman Filter (EKF) is the most popular algorithm for attitude estimation in UAVs \cite{trawny2005indirect, ko2016sine, jing2017attitude}.
EKF estimation is very accurate and widely used in practical scenarios, particularly on open-source autopilot softwares like Ardupilot and PX4.
However, the Extended Kalman filter has a few limitations of its own.
Firstly, it can be complicated to implement, which is reflected by the numerous proposed methods to improve efficiency \cite{madgwick2010efficient, simon2006optimal}.
Secondly, determining Kalman gain after every time interval requires two steps: prediction and update, thus requiring more computations to calculate mean and covariance, and larger memory to store the results.
Finally, the EKF scheme also assumes Gaussian uncertainty in its modeling, which is reasonable for uncertainty propagation over short intervals of time, but requires the algorithm to run at a higher rate resulting in larger processor usage.
These aspects make it difficult to implement EKF in low power microprocessors.

In this work, we propose a novel estimation technique called extended $\mathcal{H}_2$ filter that can overcome the limitations of the EKF, specifically with respect to computational speed, memory usage, and root mean squared error.
Since we have previously demonstrated the technique using Euler angles \cite{9147415}, we focus exclusively on the quaternion representation in this paper.
The unique nature of the quaternion vector prohibits a direct application of the general estimation algorithm.
An error dynamics model of sensors is derived to estimate the error quaternion.
The $\mathcal{H}_2$ optimal filter gain is designed offline about a nominal operating point by solving a convex optimization problem, and the  filter dynamics is implemented using the nonlinear system dynamics.
This implementation of this $\mathcal{H}_2$ optimal estimator is referred to as the \textit{extended} $\mathcal{H}_2$ estimator.
Attitude estimation using quaternions is not a novel idea \cite{choukroun2011spacecraft}, but it has not been investigated in the context of sensor fusion for drones using MEMS sensors.
The proposed technique is tested on four cases corresponding to long time-scale motion, fast time-scale motion, transition from hover to forward flight for VTOL aircrafts, and an entire flight cycle (from take-off to landing).
Its results are compared against that of the EKF in terms of the aforementioned performance metrics.

The paper is organized as follows. In section \S \ref{Section:SensorMeasurement }, we present
details of the measurement model for the three sensors.
In section \S \ref{Section: Structure}, we briefly outline the estimation framework.
Next, we describe the conventional $\mathcal{H}_2$ optimal attitude estimation algorithm.
We elaborate upon our proposal for the extended $\mathcal{H}_2$ attitude estimation algorithm using quaternions in section \S \ref{Section:EH2}.
Proceeding further, we apply the proposed filter on an illustrative example using a commercially available sensor and compare our performance with that of the popularly used EKF-based estimator.
The paper concludes with a few final remarks and potential directions for further investigation.

\section{Sensor Measurement Model} \label{Section:SensorMeasurement }

The Newton Euler approach is used to present the dynamics of a rigid body system \cite{goldstein2002classical}.
The dynamics of the vehicle is expressed in the inertial frame ($I$) and Body frame ($B$) \cite{carrillo2013modeling}.

In the body frame, the $B_x$ and $B_y$ axes point towards the heading of vehicle and starboard respectively while the $B_z$ axis points downwards.
The axes of the inertial frame are collectively referred to as NED, short for North, East, and Down.

Data collected from MEMS sensors on small UAV systems tends to be corrupted by noise and bias.
Following the Allan variance analysis \cite{allen1993performance,el2007analysis,lam2003gyro}, sensor models are described in this section.
During estimation, gyroscope model will be used for state prediction, whereas the accelerometer and magnetometer models will be used for state update.
\subsection{Gyroscope Model}
Gyroscope sensor measurements are modeled as:
\begin{subequations}
\begin{align}
& ^B\omegab = \omegab_m - \vo{b} - \vo{n}_\omegab,\\
& \dot{\vo{b}} = \vo{n_b},
\end{align}\label{gyro_model}
\end{subequations}
with the true angular rate of the body{$^B \omegab$}, angular rate measured by the gyroscope $\omegab_m$,
bias of gyroscope $\vo{b}$, gyroscope sensor noise $\vo{n}_\omegab$, and gyroscope bias noise $\vo{n_b}$.
In this paper, the gyroscope bias is non-static and modeled as a random walk process.

\subsection{Accelerometer Model}
Accelerometer sensor measurements can be formulated as:
\begin{align} \label{eqn:ACC_Meas}
{}^B{\vo{a}} &= \vo{a}_m - \vo{n}_a
\end{align}
with the true sum of the gravity and external acceleration of the body $^B \vo{a}$, sum of the gravity and external acceleration of the body measured by accelerometer  $^B \vo{a}$, and accelerometer sensor noise $\vo{n}_a$.
The external acceleration of the vehicle is derived from position estimation and subtracted from $^B \vo{a}$ to obtain acceleration due to gravity.

\subsection{Magnetometer Model}
Magnetometer sensor measurements can be formulated as:
\begin{align}\label{eqn:MAG_Meas}
^B{\vo{m}} &= \vo{m}_m - \vo{n}_m
\end{align}
with the true magnetic field $^B \vo{m}$,  the magnetic field measured by the magnetometer $\vo{m}_m$, and magnetometer sensor noise $\vo{n}_m$.

\section{Attitude Estimation System Structure for Sensor Fusion}
\label{Section: Structure}
The system equation is derived from the gyroscope dynamics and the measurement equations are composed of accelerometer and magnetometer models. Additionally, error dynamics is derived for estimation when using quaternions.

\subsection{System Dynamics}
\subsubsection{System Equations with Quaternion}
Using the definition of the quaternion derivative \cite{shuster1993survey} and the gyroscope sensor model (\ref{gyro_model}), the system of differential equations is obtained as:
\begin{subequations}\label{eqn:gryo_quat}
\begin{align}
    ^B_I\dot{\Bar{\vo{q}}} &=\frac{1}{2} \vo{\Omega}(\omegab) ^B_I \Bar{\vo{q}},\\
    \dot{\vo{b}} &= \vo{n}_\omegab
\end{align}
\end{subequations}
where:
\begin{align}\label{quaternian Omega}
{
\vo{\Omega}(\omegab) :=
  \begin{bmatrix}
    0   & \omegab_z  & -\omegab_y & \omegab_x \\
   -\omegab_z & 0    & \omegab_x  & \omegab_y \\
    \omegab_y & -\omegab_x & 0    & \omegab_z \\
   -\omegab_x & -\omegab_y & -\omegab_z & 0
   \end{bmatrix},}\
   \Bar{\vo{q}}:= \begin{pmatrix} \vo{q} \\ q_4 \end{pmatrix}= \begin{pmatrix} q_1 \\ q_2 \\ q_3 \\ q_4 \end{pmatrix}
\end{align}
with $\omegab$ is the angular velocity of the vehicle in body coordinates.

\subsubsection{Measurement Equations with Quaternion}

The attitude of the vehicle in quaternions is recovered from the DCM using quaternions by comparing sensor values in the body frame against  gravity or Earth's magnetic field vector in the inertial frame.

The relationship between the gravity vector ${}^I \vo{g}$ in the inertial frame and the acceleration vector ${}^B \vo{a}$ in body frame from the accelerometer measurement (\ref{eqn:ACC_Meas}) can be formulated with quaternions as:
\begin{align}\label{Acc_model_quat}
   {}^B \vo{a} = {\vo{C}^{B}_I}_{acc} (\Bar{\vo{q}}) \hspace{0.1cm} {}^I\vo{g}
\end{align}
with ${\vo{C}^B_I}_{acc}$ is the DCM from inertial frame ($I$) to body frame ($B$).

The relationship between the Earth's magnetic vector ${}^I \vo{h}$ and the local magnetic vector ${}^B \vo{m}$ from  magnetometer measurement (\ref{eqn:MAG_Meas}) can be expressed as:
\begin{align}
   {}^B \vo{m} = {\vo{C}^B_I}_{mag} (\Bar{\vo{q}}) \hspace{0.1cm} {}^I \vo{h} \label{mag_model_quat}
\end{align}
Therefore, the final measurement equations combining the two sensor models in (\ref{eqn:ACC_Meas}) and (\ref{eqn:MAG_Meas}) are formulated as:
\begin{align}\label{eqn:meas_quat}
    \begin{pmatrix}^B\vo{a}_m \\ ^B \vo{m}_m\end{pmatrix} =
    \begin{bmatrix}
       \vo{C}_{acc}(\vo{\Bar{q}})& \vo{C}_{mag}(\vo{\Bar{q}})
    \end{bmatrix}
    \begin{pmatrix}\vo{g}\\\vo{h}\end{pmatrix} + \begin{pmatrix}\vo{n_a}\\\vo{n_m}\end{pmatrix}
\end{align}
and, the sensor noise covariance matrix for $\begin{pmatrix}\vo{n_a}\\\vo{n_m}\end{pmatrix}$ is given by:

\begin{gather}
\vo{R} =
  \begin{bmatrix}
   \vo{N}_a & \vo{0}_{3\times3}\\
   \vo{0}_{3\times3} & \vo{N}_m
   \end{bmatrix}
=
  \begin{bmatrix}
  n^2_a \vo{I}_{3\times3} & \vo{0}_{3\times3}\\
   \vo{0}_{3\times3} & n^2_m \vo{I}_{3\times3}
   \end{bmatrix}
\end{gather}

\subsection{Error Dynamics}

\subsubsection{Error System Equations with Quaternion}

Instead of using the arithmetic difference between quaternion and quaternion estimate to define the error, we will introduce the error quaternion $\delta \hat{\vo{q}}$; a small rotation between the estimated and the true orientation of the body frame of reference.

This error calculation is expressed as a  multiplication in quaternion algebra \cite{kuipers1999quaternions} is:
\begin{align}
    {}^{B}_{I} \Bar{\vo{q}} &= {}^B_{\hat{B}} \delta \Bar{\vo{q}} \otimes \vspace{0.2cm} {}^{\hat{B}}_{I} \hat{\Bar{\vo{q}}} \label{eqn:Error_quat}\\
    {}^{B}_{\hat{B}} \delta \Bar{\vo{q}} &= {}^{B}_{I} \Bar{\vo{q}} \otimes \vspace{0.2cm} {}^{\hat{B}}_{I} \hat{\Bar{\vo{q}}}^{-1}
\end{align}
Note that here, $\otimes$ is used to indicate a product of two terms in quaternion algebra, and is NOT the Kronecker product.

We can apply the small angle approximation to $\delta \Bar{\vo{q}}$ assuming the rotation associated with the error quaternion is very small. Consequently, the attitude error angle vector $\delta  \vo{\theta}$ is  calculated as:
\begin{align}\label{eqn:small_anlge}
    \delta \Bar{\vo{q}} =
    \begin{pmatrix}
    \delta \vo{q} \\
    \delta q_4
    \end{pmatrix} =
    \begin{pmatrix}
    \delta \vo{\hat{k}} \sin{(\delta {\theta} /2)}\\
    \cos{(\delta \theta /2)}
    \end{pmatrix} \approx
    \begin{pmatrix}
    \frac{1}{2} \delta \vo{\theta}\\
    1
    \end{pmatrix}
\end{align}
Here, error angle vector $\delta  \vo{\theta} \in \mathbb{R}^3$ will be used together with the bias error in the error state vector, which is defined as:
\begin{align}
    \Delta \vo{b} = \vo{b} - \hat{\vo{b}}
\end{align}
From  \cite{trawny2005indirect}, the definition of the error quaternion (\ref{eqn:Error_quat}) results in the following set of error system equations:
\begin{subequations}
\begin{align}
    \delta \dot{\vo{\theta}} &= -[\hat{\vo{\omegab}} \times] \delta \vo{\theta} -\Delta \vo{b} -\vo{n_w}\\
    \Delta \dot{\vo{b}} &= \dot{\vo{b}}  - \dot{\hat{\vo{b}}} =  \vo{n_b}
\end{align}
\end{subequations}
or, in state space form as:
\begin{align}\label{eqn:Gyro_M_quat}
\begin{pmatrix} \dot{\delta \vo{\theta}} \\ \dot{\Delta \vo{b}} \end{pmatrix}
 = \begin{bmatrix} -\hat{\vo{\omegab}}\times & -I_{3\times3} \\ \vo{0}_{3\times3} & \vo{0}_{3\times3}\end{bmatrix}\begin{pmatrix}\delta \vo{\theta}\\\Delta \vo{b} \end{pmatrix} +
 \begin{bmatrix}-I_{3\times3} & \vo{0}_{3\times3} \\ \vo{0}_{3\times3} & I_{3\times3}\end{bmatrix}\begin{pmatrix}\vo{n}_w\\\vo{n_b}\end{pmatrix},
\end{align}
The process noise covariance matrix for $\begin{pmatrix}\vo{n_w}\\\vo{n_b}\end{pmatrix}$ is given by:
\begin{gather}
\vo{Q} =
  \begin{bmatrix}
   \vo{N}_w & \vo{0}_{3\times3}\\
   \vo{0}_{3\times3} & \vo{N}_b
   \end{bmatrix}
=
  \begin{bmatrix}
  n^2_w\vo{I}_{3\times3} & \vo{0}_{3\times3}\\
   \vo{0}_{3\times3} & n^2_b \vo{I}_{3\times3}
   \end{bmatrix}
\end{gather}
\subsubsection{Error Measurement Equations with Quaternion}

Error dynamics is used to estimate the error quaternion $\delta \hat{\vo{q}}$, defined previously in the error system equation. Recall that the measurement equations with quaternion (\ref{eqn:meas_quat}) are in terms of the DCM and hence, we obtain the error in the estimate in the following manner.

\begin{align}
    \Tilde{\vo{y}} &= \vo{y} -\hat{\vo{y}} = ({\vo{C}^{B}_{I}}(\Bar{\vo{q}})  - \vo{C}^{\hat{B}}_{I}(\hat{\Bar{\vo{q}}})) \ ^I \begin{bmatrix}
       \vo{g} \\ \vo{h}\end{bmatrix}
 \end{align}
where, $\vo{y}$ is actual measurement and $\hat{\vo{y}}$ is estimated using $\hat{\Bar{\vo{q}}}$ from the result of original system equations (\ref{eqn:gryo_quat}).
With this equation, we can derive the final error measurement equations with accelerometer and magnetometer model.
These can be written using the DCM as:
\begin{align}
    \Tilde{\vo{y}} &= \vo{C}^{\hat{B}}_I (\hat{\vo{q}})\begin{bmatrix}\vo{g}\\\vo{h}\end{bmatrix} + \vo{D_w} \vo{w}(t)
\end{align}
where,
\begin{gather*}
\vo{C}^B_I(\hat{\vo{q}})
   =
    \begin{bmatrix}
       \vo{C}_{acc}(\hat{\Bar{\vo{q}}})& \vo{C}_{mag}(\hat{\vo{q}})
    \end{bmatrix},\hspace{0.3cm}
\vo{D_w} =
   \begin{bmatrix}
        \vo{I}_{3\times3} & \vo{0}_{3\times3}\\
        \vo{0}_{3\times3} & \vo{I}_{3\times3}
  \end{bmatrix}
\end{gather*}

\section{$\mathcal{H}_2$ Optimal Estimation}
We next present very briefly, the necessary background for $\mathcal{H}_2$ optimal estimation method for linear systems. We consider the following linear system,
\begin{subequations}\label{system_CT}
\begin{align}
\Dot{\vo{x}}(t) &=  \vo{A} \vo{x}(t) + \vo{B}_u \vo{u}(t) + \vo{B}_w \vo{w}(t)\\
\vo{y}(t) &= \vo{C}_y \vo{x}(t) + \vo{D}_u \vo{u}(t)+ \vo{D}_w \vo{w}(t)\\
\vo{z}(t) &= \vo{C}_z \vo{x}(t)
\end{align}
\end{subequations}
with $\vo{x} \in \mathbb{R}^{n}$, $\vo{y} \in \mathbb{R}^{l}$, $\vo{z} \in \mathbb{R}^{m}$ are respectively the state vector, the measured output vector, and the output vector of interest. Variables $\vo{w} \in \mathbb{R}^{p}$ and $\vo{u} \in \mathbb{R}^{r}$ are the disturbance and the control vectors, respectively.

With the above defined system, the $\mathcal{H}_2$ state estimator has the following form,
\begin{subequations} \label{estimation_CT}
\begin{align}
    \Dot{\hat{\vo{x}}}(t) &= \vo{A} \hat{\vo{x}}(t) + \vo{B}_u \vo{u}(t) + \vo{L} (\vo{C}_y \hat{\vo{x}}(t) + \vo{D}_u \vo{u}(t) - \vo{y}(t)),\\
    \hat{\vo{z}}(t) &= \vo{C}_z \hat{\vo{x}}(t),
\end{align}
\end{subequations}
where $\hat{\vo{x}}$ is the state estimate, $\vo{L}$ is the estimator gain, and $\hat{\vo{z}}(t)$ is the estimate of the output of interest. The error equations are then given by:
\begin{subequations} \label{estimator_Linear}
    \begin{align}
    \hat{\Dot{\vo{e}}}(t) &= (\vo{A}+\vo{L} \vo{C}_y) \hat{\vo{e}}(t) + (\vo{B}_w+\vo{L} \vo{D}_w)\vo{w}(t)\\
    \Tilde{\vo{z}}(t) &= \vo{C}_z \hat{\vo{e}}(t)
    \end{align}
\end{subequations}

The problem of $\mathcal{H}_2$ state estimation design can then be stated as: Given a system (\ref{estimator_Linear}) and a positive scalar $\gamma$, find a matrix $\vo{L}$ such that,
\begin{align}\label{min1}
\|\vo{G}_{\Tilde{z}w}(s)\|_{2} < \gamma.
\end{align}
where the transfer function $\vo{G}_{\Tilde{z}w}(s)$ of the system is:
\begin{align}\label{errorTF}
\vo{G}_{\Tilde{z} w}(s) = \vo{C}_z [s \vo{I} -( \vo{A} + \vo{L} \vo{C}_y)]^{-1} (\vo{B}_w + \vo{L} \vo{D}_w) .
\end{align}

The optimization formulation to obtain $\vo{L}$ is given by:\\[2mm]
\textbf{Theorem ($\mathcal{H}_2$ Optimal Estimation)}  {\cite[p.~296]{duan2013lmis},\cite{apkarian2001continuous}} : The following two statements are equivalent:
\begin{enumerate}
\item A solution $\vo{L}$ to the $\mathcal{H}_2$ state estimator exists.
\item $\exists$ a matrix $\vo{W}$, a symmetric matrix $\vo{Q}$, and a symmetric matrix $\vo{X}$ such that:
\end{enumerate}
        \begin{align}\label{CT_LMI}
        \begin{bmatrix}\nonumber
        \vo{XA}+\vo{W} \vo{C}_y +(\vo{XA}+\vo{W} \vo{C}_y)^T &  \vo{X} \vo{B}_w+\vo{W} \vo{D}_w\\
        \vo{*}      & -\vo{I}
        \end{bmatrix}
        < 0\\
        \begin{bmatrix}\nonumber
        \vo{-Q}  &  \vo{C}_z\\
        \vo{*} & \vo{-X}
        \end{bmatrix}
        < 0\\
         \textbf{trace}(\vo{Q}) < \gamma^2
        \end{align}
{
The minimal attenuation level $\gamma$ is determined by solving the following optimization problem
\begin{align}
\min_{\vo{W}, \vo{X}, \vo{Q}} \gamma \quad \text{subject to (\ref{CT_LMI}). }  \nonumber
\end{align}
To solve this optimized solution, we used CVX \cite{grant2009cvx} and MATLAB tool box \cite{gu2005robust}.
}
The $\mathcal{H}_2$ optimal estimator gain is recovered by $\vo{L} = \vo{X}^{-1} \vo{W}$. This optimal gain ensures that:
\begin{align}
\vo{e}(t) = \vo{x}(t) - \hat{\vo{x}} \rightarrow 0, \hspace{0.5cm} \text{ as } t \rightarrow \infty,
\end{align}
In other words, $\hat{\vo{x}}(t)$ is an asymptotic estimate of $\vo{x}(t)$.

\section{Extended $\mathcal{H}_2$ Optimal Estimation} \label{Section:EH2}

\begin{figure}[h!]
\centering
\includegraphics[width=0.95\textwidth]{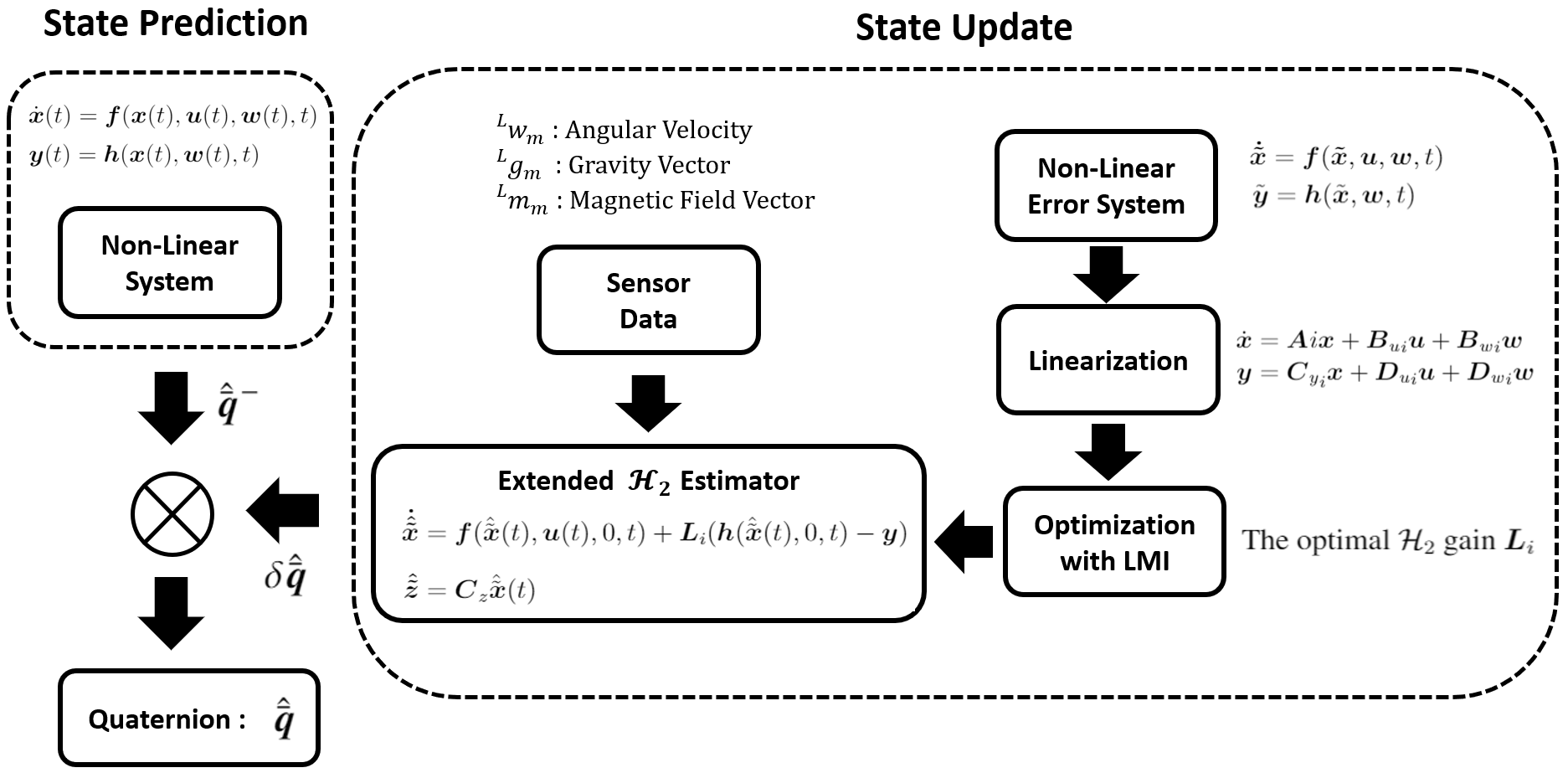}
\caption{Quaternion estimation algorithm}
\label{Fig:algorithm_quaternion}
\end{figure}

The process for extended $\mathcal{H}_{2}$ estimation using quaternions is not too different from that developed for using Euler angles \cite{9147415}.
The major distinction arises when solving for the error in the error dynamics model, since quaternion vector algebra requires special attention.
The error state between the true quaternion and the predicted quaternion estimate expressed as $\delta \hat{\vo{q}}$ is used for estimation, as stated in (\ref{eqn:small_anlge}).
Another distinguishing characteristic is the introduction of gain scheduling to adequately cover all possible orientations of the vehicle.
We obtain these gains as a function of the linearization points, and apply them during the nonlinear update step.
The proposed extended $\mathcal{H}_2$ estimation framework for quaternion is outlined in Fig.\ref{Fig:algorithm_quaternion}. Like any estimation procedure, this is essentially composed of two steps: state prediction and state update.

\subsection{State Prediction}
To express the dynamics of the quaternion system with state space notation, we define a 7-element state vector.
\begin{align}
    \vo{x}(t) =
    \begin{pmatrix}
        \Bar{\vo{q}}\\
        \vo{b}
    \end{pmatrix}
\end{align}
using the definition of the quaternion derivative \cite{kuipers1999quaternions}.
The non-linear system of gyroscope dynamics with quaternion (\ref{eqn:gryo_quat}) can be rewritten with states as:
\begin{align}\label{eqn:sys_quat}
\Dot{\vo{x}} &= \vo{f}(\vo{x},\vo{u},\vo{w},t),
\end{align}
where $\vo{u}(t) := \omegab_m(t)$ and $\vo{w}(t):=
        \left [\vo{n}_\omegab(t) \quad
        \vo{n}_b(t) \right]$.

Taking the expectation of the above equation (\ref{eqn:sys_quat}) derived from the quaternian gyroscope dynamics (\ref{eqn:gryo_quat}), we can get prediction equations as:
\begin{align}\label{eqn:sys_quat_est}
\Dot{\hat{\vo{x}}} &= \vo{f}(\hat{\vo{x}},\vo{u},\vo{w},t),
\end{align}
i.e.,
\begin{subequations}\label{eqn:gryo_quat_estimation}
\begin{align}
    ^B_I\dot{\hat{\Bar{\vo{q}}}}^- &=\frac{1}{2} \vo{\Omega}(\omegab_m - \hat{\vo{b}}) ^B_I \hat{\Bar{\vo{q}}},\\
    \dot{\hat{\vo{b}}} &= \vo{0}_{3 \times 1}
\end{align}
\end{subequations}
With this dynamic equation, we can get an predicted estimate of $\hat{\Bar{\vo{q}}}^-$ in Fig.\ref{Fig:algorithm_quaternion}, which will be updated
with error dynamics defined appropriately for quaternions.

\subsection{State Update}
\subsubsection{Error State Equations:}
We defined the error between the true quaternion  and quaternion estimate as $\delta \hat{q}$ and applied small angle approximation as $\delta \vo{\theta}$ in (\ref{eqn:small_anlge}) in previous section. We can now define a 6-element state vector for the estimation of the error dynamics as:
\begin{align}
    \Tilde{x} =
    \begin{pmatrix}
    \delta \vo{\theta} \\
    \Delta \vo{b}
    \end{pmatrix}
\end{align}
With this state, error system equation with quaternion (\ref{eqn:Gyro_M_quat}) can be rewritten as a nonlinear system as:
\begin{align}
\vo{\dot{\Tilde{x}}} = \vo{f}(\vo{\Tilde{x}},\vo{u},\vo{w},t). \label{eqn:nldyn_quat}
\end{align}
where $\vo{u}(t) := \omegab_m(t)$ and $\vo{w}(t):=
        \left [\vo{n}_\omegab(t) \quad
        \vo{n}_b(t) \right]$.

A linear approximation is implemented at the chosen nominal operating point $(\vo{x}_0,\vo{u}_0, \vo{w}_0) = \vo{0}$.
The linear equations are:
\begin{align}
     \Dot{\vo{x}}(t) &\approx \vo{A} \vo{x}(t) + \vo{B_w} \vo{w}(t)
 \end{align}
i.e.,
\begin{align}\label{eqn:Gyro_M_quat_Lin}
\begin{pmatrix} \dot{\delta \vo{\theta}} \\ \dot{\Delta \vo{b}} \end{pmatrix}
 = \begin{bmatrix} -|\hat{\vo{\omegab}}\times| & -I_{3\times3} \\ \vo{0}_{3\times3} & \vo{0}_{3\times3}\end{bmatrix}\begin{pmatrix}\delta \vo{\theta}\\\Delta \vo{b} \end{pmatrix} +
 \begin{bmatrix}-I_{3\times3} & \vo{0}_{3\times3} \\ \vo{0}_{3\times3} & I_{3\times3}\end{bmatrix}\begin{pmatrix}\vo{n}_w\\\vo{n_b}\end{pmatrix},
\end{align}
where $\hat{\vo{\omegab}} := \omegab_m - \hat{\vo{b}}$.
\subsubsection{\textbf{Error Measurement Equation:}}
The measurement equation of the error system with accelerometer and magnetometer can be written as the following nonlinear output equation:
\begin{align}\label{eqn:error_meas}
    \Tilde{\vo{y}} &= \vo{h}(\Tilde{\vo{x}},\vo{w},t) = \vo{C}^{\hat{B}}_I (\hat{\vo{q}})\begin{bmatrix}\vo{g}\\\vo{h}\end{bmatrix} + \vo{D_w} \vo{w}(t)
\end{align}
where,
\begin{gather*}
\vo{C}^B_I(\hat{\vo{q}})
   =
    \begin{bmatrix}
       \vo{C}_{acc}(\hat{\Bar{\vo{q}}})& \vo{C}_{mag}(\hat{\vo{q}})
    \end{bmatrix},\hspace{0.3cm}
\vo{D_w} =
   \begin{bmatrix}
        \vo{I}_{3x3} & \vo{0}_{3x3}\\
        \vo{0}_{3x3} & \vo{I}_{3x3}
  \end{bmatrix}
\end{gather*}
{We apply a linear approximation to (\ref{eqn:error_meas}) about eight points, each of which is the combination of a point from each axis covering a range of $[\frac{\pi}{2} $ $ \frac{\pi}{2} ]$ and  $[\frac{\pi}{2}$ \vspace{5mm} $-\frac{\pi}{2}]$, i.e., the following eight points: $(0,0,0)$, $(0,0,\pi)$, $(0,\pi,0)$, $(\pi,0,0)$, $(0,\pi,\pi)$, $(\pi,\pi,0)$, $(\pi,0,\pi)$, and $(\pi,\pi,\pi)$.}
The linear system, about a nominal operating point, is therefore:
\begin{subequations}\label{Lin_systems_quat}
\begin{align}
\Dot{\vo{x}} &=  \vo{A}_i  \vo{x} + {\vo{B}_u}_i  \vo{u} + {\vo{B}_w}_i  \vo{w}\\
\vo{y} &= {\vo{C}_y}_i  \vo{x} + {\vo{D}_u}_i  \vo{u}+ {\vo{D}_w}_i  \vo{w}
\end{align}
\end{subequations}
i.e.,
\begin{gather} \label{eqn:Lin_systems_quat_mtx}
\vo{y}(t) =
    \begin{bmatrix}
       [\vo{g} {\times}]_i & \vo{0}_{3x3} \\
       [\vo{h} {\times}]_i & \vo{0}_{3x3}\\
    \end{bmatrix}
    \vo{x}(t)
+    \begin{bmatrix}
        \vo{I}_{3x3} & \vo{0}_{3x3}\\
        \vo{0}_{3x3} & \vo{I}_{3x3}
  \end{bmatrix}
  \vo{w}(t)
\end{gather}
where $\vo{g} {\times}$ and $\vo{h} {\times}$ are the skew-symmetric matrix forms of the vectors $\vo{g}$ and $\vo{h}$ respectively and the
signs vary with the linearization point in consideration.
The optimal $\mathcal{H}_2$ gain $\vo{L}_i$ can then be determined by solving the optimization problem in (\ref{CT_LMI}), where the subscript $i (=1\sim8)$ is used to indicate that it is determined about a nominal operating point.
Once the gain $\vo{L}_i$ is determined, it is used to implement the $\mathcal{H}_2$ filter for the nonlinear system. Note that here, gain scheduling is applied, depending on the system states.
This is because signs of $\vo{g}$ and $\vo{h}$ in the measurement equation (\ref{eqn:Lin_systems_quat_mtx}) are inverted while the vehicle is rotating.
The filter dynamics and output equation, for the extended $\mathcal{H}_2$ filter, are given by:
\begin{subequations}\label{estimator_q}
    \begin{align}
     \vo{\Dot{\Hat{\Tilde{x}}}} &= \vo{f}(\hat{\Tilde{\vo{x}}}(t),\vo{u}(t),0,t) +  \vo{L}_i(\vo{h}(\Hat{\Tilde{\vo{x}}}(t),0,t) - \vo{y})\\
    \vo{\hat{\Tilde{z}}} &= \vo{C}_z \hat{\Tilde{\vo{x}}}(t)
    \end{align}
\end{subequations}

The error state of quaternion ($\hat{\delta \vo{\theta}}$) from the estimation result of previous step is recovered from equation (\ref{eqn:small_anlge}) as:
\begin{align}
\begin{bmatrix}
\delta  \hat{\vo{q}}\\ \Delta  \hat{\vo{b}}
\end{bmatrix}
=
\begin{bmatrix}
\hat{\delta \vo{\theta}}/2 \\ \Delta  \hat{\vo{b}}
\end{bmatrix}
\end{align}

We need to ensure that the unit norm constraint of the estimated quaternion is not violated. The full quaternion satisfying the unit norm constraint can be recovered as:
\begin{align}
    \delta \hat{\Bar{\vo{q}}} =
    \begin{bmatrix}
        \delta {\hat{\vo{q}}} \\ \sqrt{1-\delta \hat{\vo{q}}^
        {T} \delta  \hat{\vo{q}}}
    \end{bmatrix}\
\text{or}\
    \delta \hat{\Bar{\vo{q}}} =
    \frac{1}{\sqrt{1-\delta \hat{\vo{q}}^
        {T} \delta  \hat{\vo{q}}}} \cdot
    \begin{bmatrix}
        \delta {\hat{\vo{q}}} \\ 1
    \end{bmatrix}\
(\text{if} \hspace{0.2cm}
\delta {\hat{\vo{q}}}^T \delta {\hat{\vo{q}}} >1 )
\end{align}

The final estimate is a quaternion multiplication of the two results from the prediction step (${\vo{\hat{\Bar{q}}}}^-$)  and the update step ($\delta \vo{\hat{\Bar{ q}}}$), i.e.,:
\begin{align}
    \hat{\Bar{\vo{q}}} &= \delta \hat{\Bar{\vo{q}}} \otimes \vspace{0.2cm} \hat{\Bar{\vo{q}}}^-
\end{align}


\section{Results}
\subsection{Simulation Set-Up}
\begin{figure}[h!]
\centering
\includegraphics[width=0.8\textwidth]{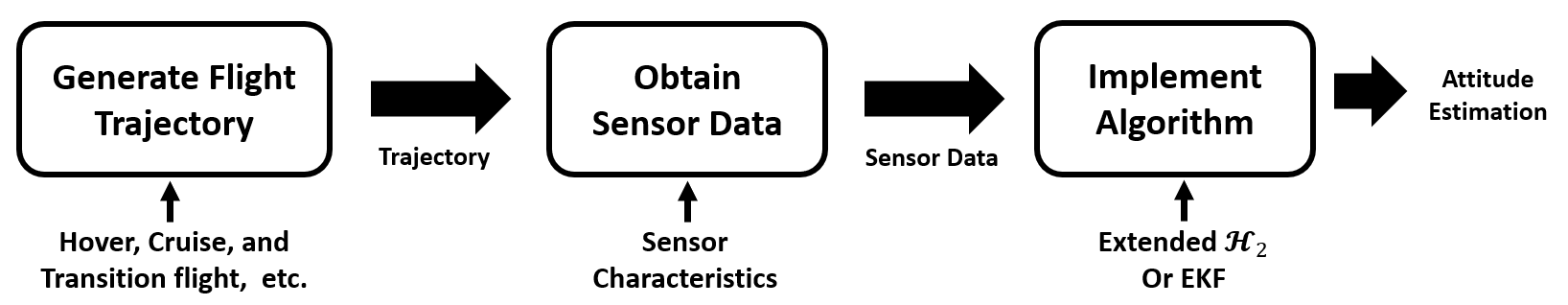}
\caption{Simulation Flow Chart}
\label{sim_chart}
\end{figure}

The implementation of the proposed extended $\mathcal{H}_2$ filter for the attitude estimation problem is tested with the MATLAB simulation environment as shown in Fig. \ref{sim_chart}.
Its performance is compared with that of the extended Kalman filter based implementation, simulated using identical sensor data.
The comparison is done in terms of accuracy and computational time under multiple scenarios of flight stages like take-off, landing, hovering, and transition flight.

\begin{figure}[h!]
\centering
  \includegraphics[width=0.8\textwidth]{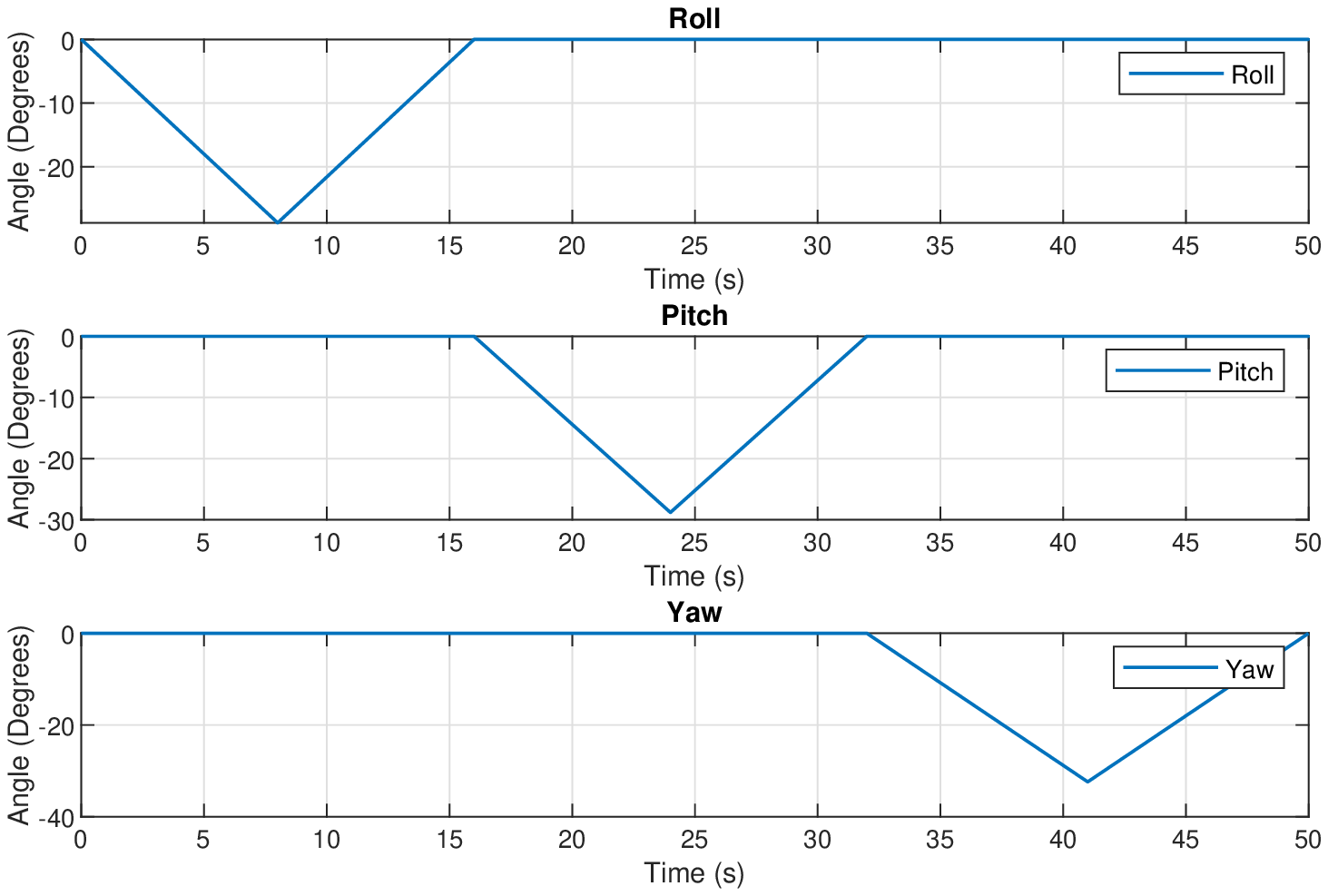}
    \caption{True trajectories imported to MATLAB for simulation.}
    \label{Tra_Low}
\end{figure}

\begin{figure}[h!]
\centering
    \includegraphics[width=0.8\textwidth]{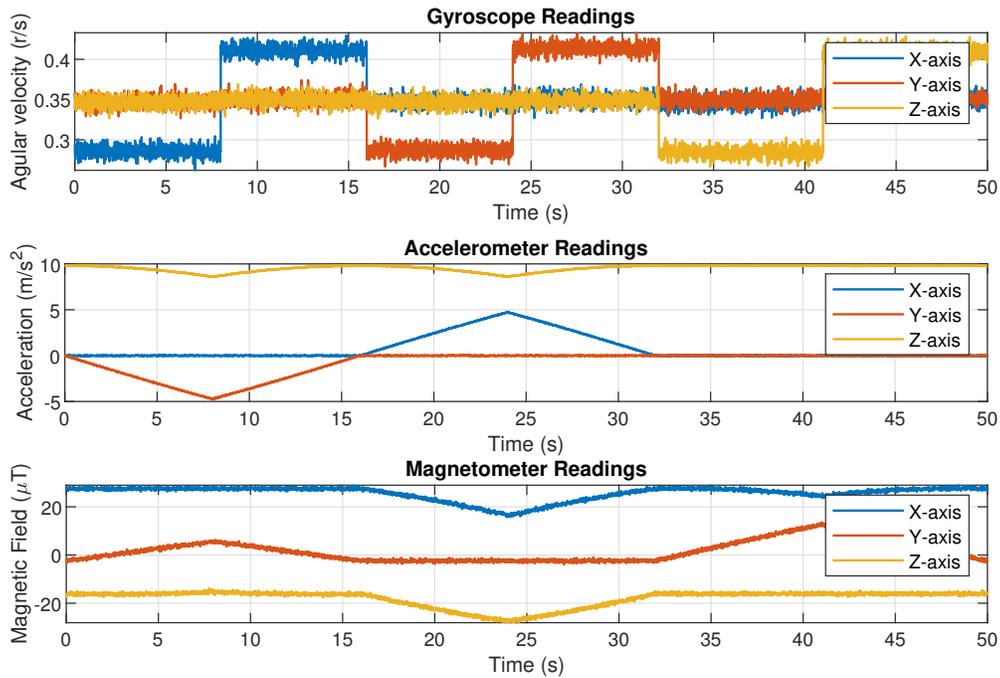}
    \caption{Sensor Data from MATLAB function \textsl{imuSensor} with MPU- 9250.}
    \label{IMU_Low}
\end{figure}

{For our experiment, we chose to simulate using the MPU 9250, an affordable commercial sensor used most popularly in the Pixhawk flight computer.
Sensor characteristics like noise levels, bias, etc. are imported to the MATLAB simulation from sensor data sheet \cite{MPU} of MPU 9250.
We use MATLAB's simulation function, \texttt{imuSensor} to generate MARG data \cite{MAT}.
This acquired raw data is shown in Fig. \ref{IMU_Low} with the trajectory shown in Fig. \ref{Tra_Low}.
The data was taken at a sample rate of 150 $Hz$.
Four scenarios covering a variety of major flight stages are used to verify the proposed extended $\mathcal{H}_2$ estimation algorithm.}

\begin{figure}[h!]
\centering
    \includegraphics[width=0.8\textwidth]{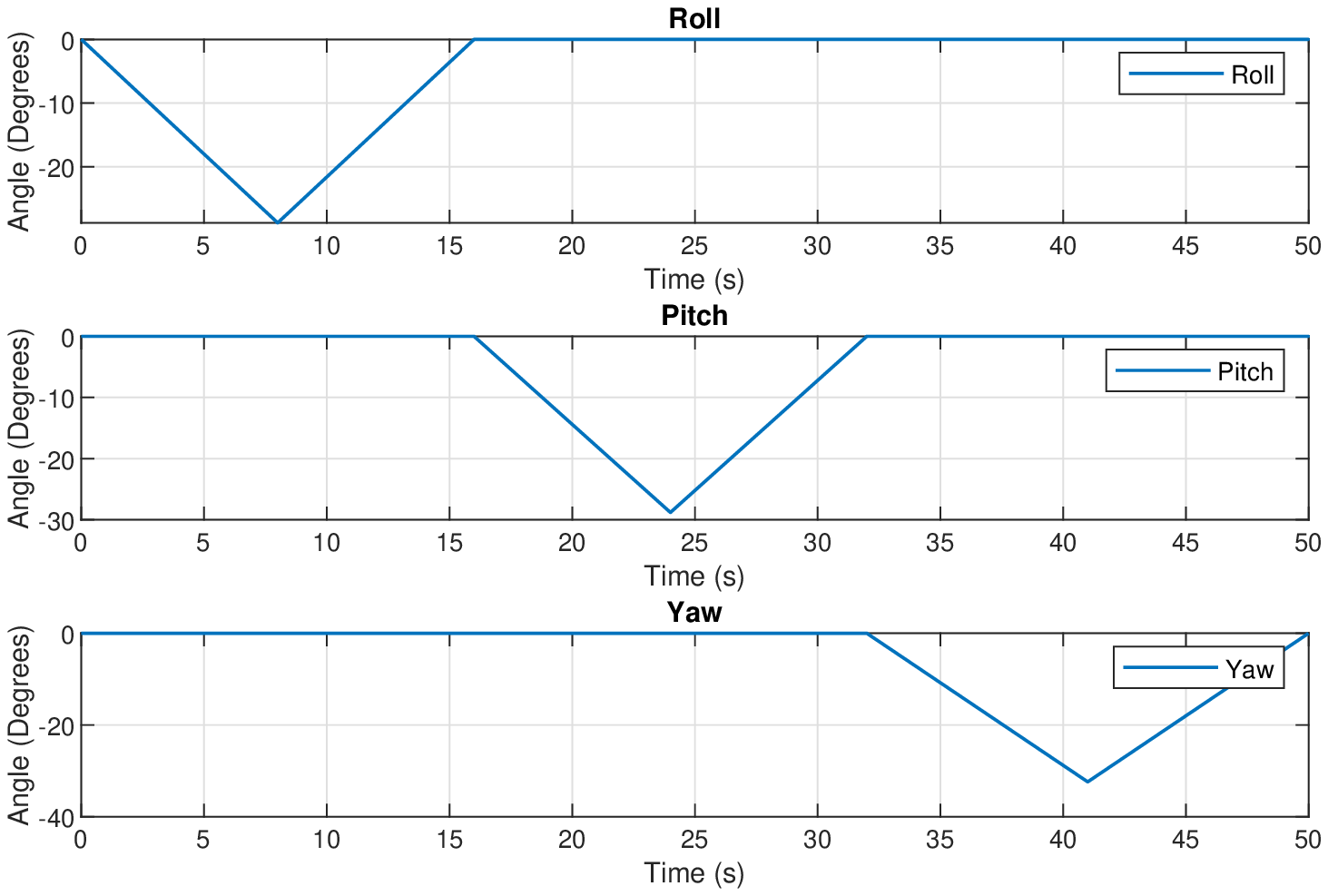}
    \caption{True trajectories for quaternion estimation in Case I.}
    \label{Tra_Low1}
\end{figure}

\textbf{Case I: Slow and Small Angular Movements} -- Here we consider angular movements $< 30^{\circ}$ about all three axes of the vehicle independently. This case broadly covers forward/backward and left/right cruise flight of popular quad-rotor based UAVs. Simulation is run for a time duration of 50 seconds and with an angular rate of $\pi/50$ rad/s. The simulated true state trajectories are shown in Fig. \ref{Tra_Low1}.

\begin{figure}[h!]
\centering
    \includegraphics[width=0.8\textwidth]{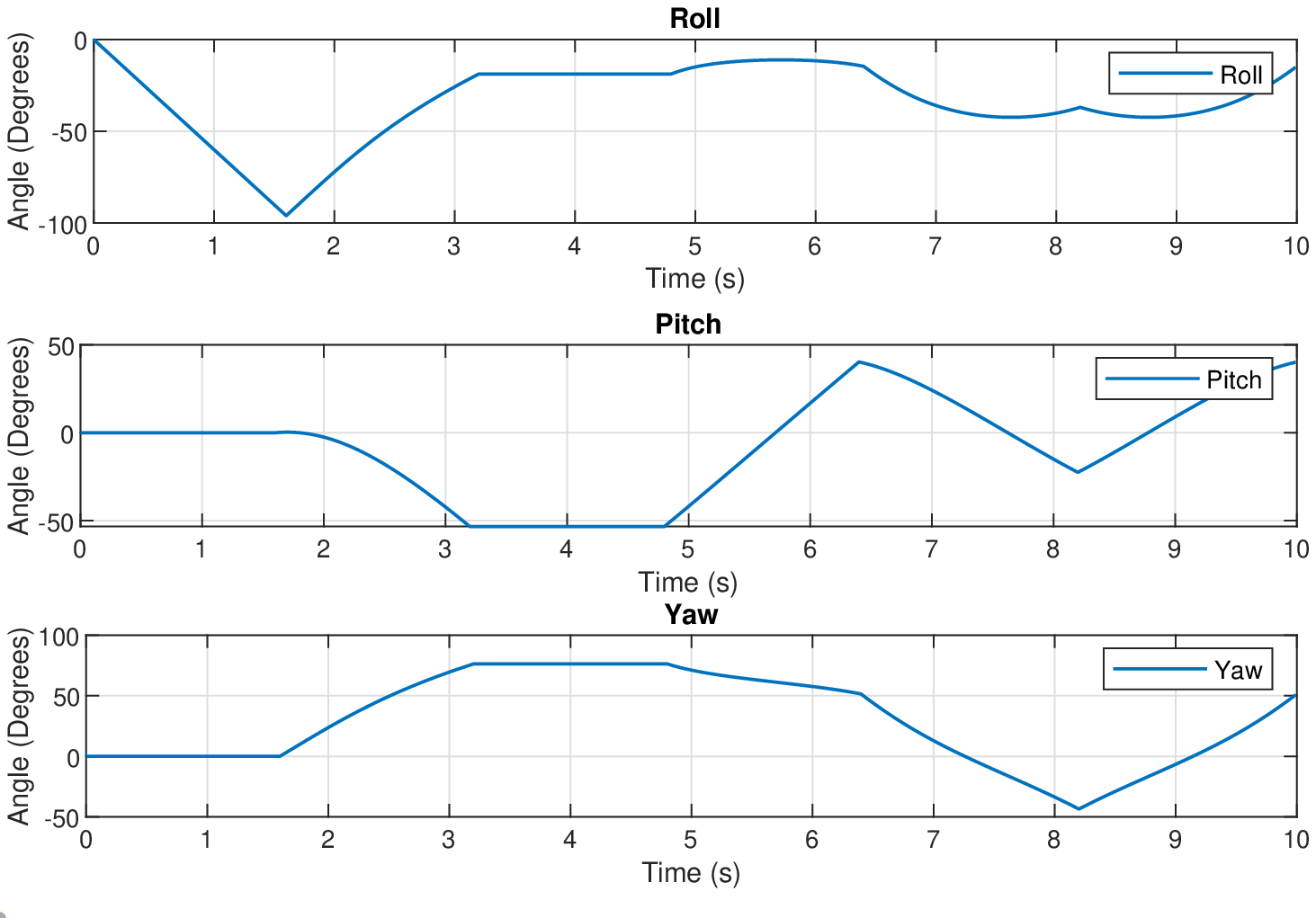}
    \caption{True trajectories for quaternion estimation in Case II.}
    \label{Tra_high}
\end{figure}

\textbf{Case II: Fast and High Angular Movements} -- Here we consider angular variation $> 30^{\circ}$  about all three axes of the vehicle simultaneously. It represents scenarios of rapid movements or motion in the presence of wind disturbance during flight or aggressive maneuvers. Simulation is run for a duration of 10 seconds and with an angular rate of $\pi/3$ rad/s. The simulated true state trajectories are shown in Fig. \ref{Tra_high}.

\begin{figure}[h!]
\centering
    \includegraphics[width=0.8\textwidth]{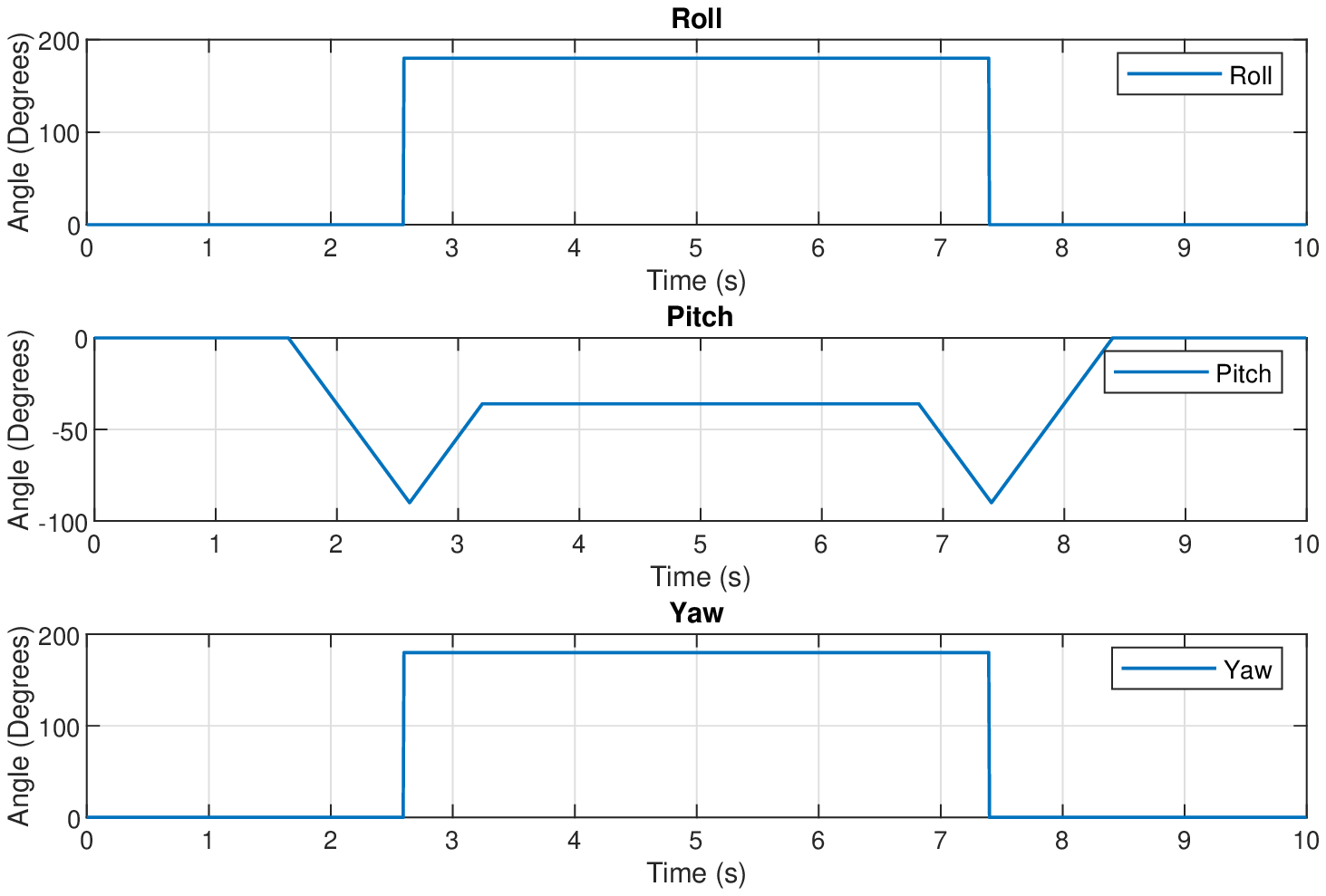}
    \caption{True trajectories for quaternion estimation in Case III.}
    \label{Fig:gimball_track}
\end{figure}

\textbf{Case III: Gimbal-lock test} -- Here we consider angular movements $> 90^{\circ}$ about pitch axis of the vehicle. This case broadly covers transition flight of the increasingly popular VTOL (Vertical Take Off and Land) UAVs. Simulation is run for a time duration of 10 seconds and with an angular rate $\pi/2$ rad/s. The simulated true state trajectories are shown in Fig. \ref{Fig:gimball_track}.

\begin{figure}[h!]
\centering
    \includegraphics[width=0.8\textwidth]{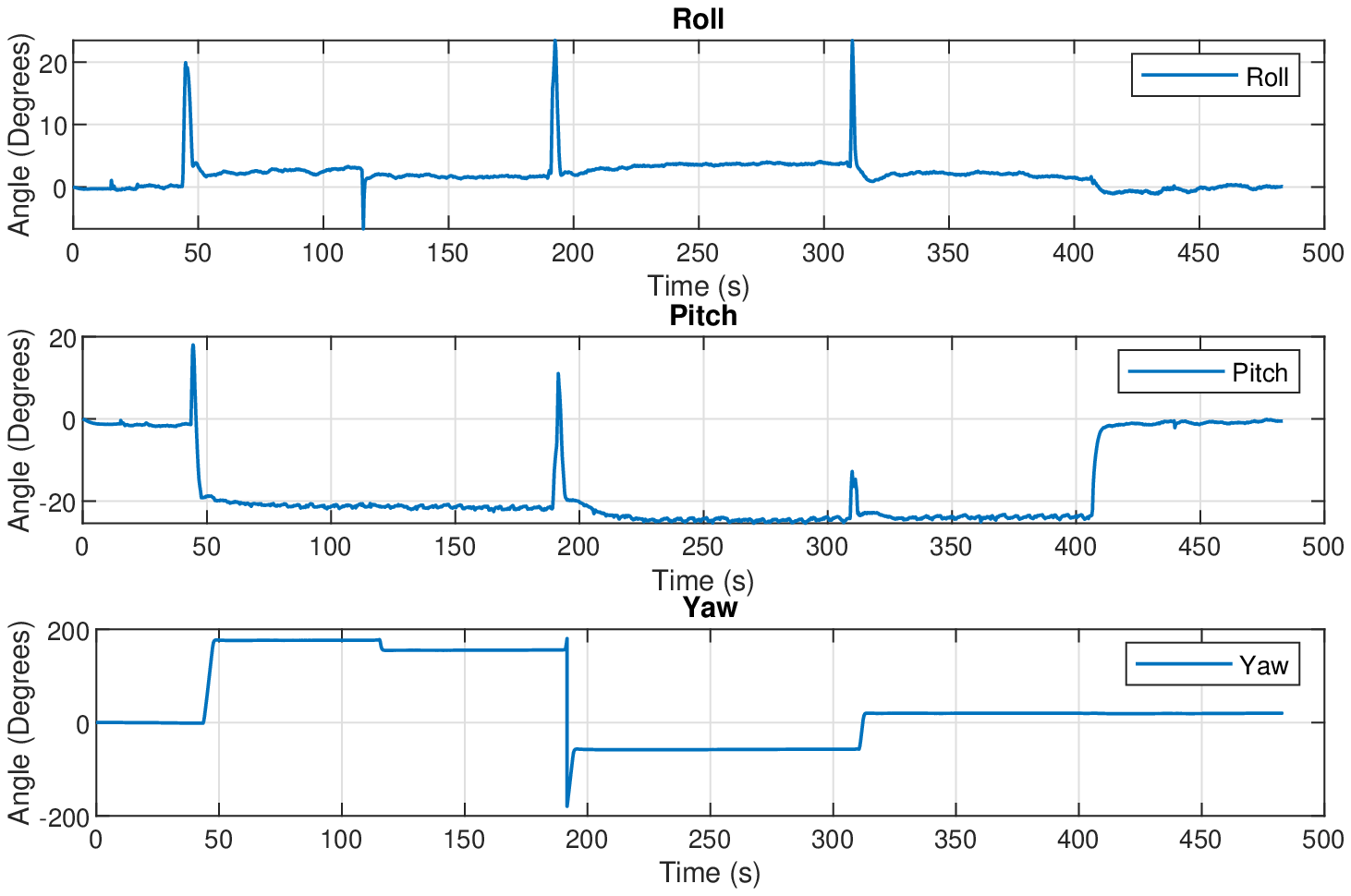}
    \caption{True trajectories for quaternion estimation in Case IV.}
    \label{Fig:quat_ref_sim}
\end{figure}

\textbf{Case IV: Movement from 3D flight simulation} -- Here we consider flight trajectory generated by SITL (Simulation In The Loop) with the Ardupilot firmware and Gazebo. This represents scenarios that include taking off, cruise to three way points and then landing. The simulated true state trajectories are shown in Fig. \ref{Fig:quat_ref_sim}.

\subsection{Simulation Results for Quaternion Estimation}
\begin{figure}[h!]
\centering
    \includegraphics[width=0.8\textwidth]{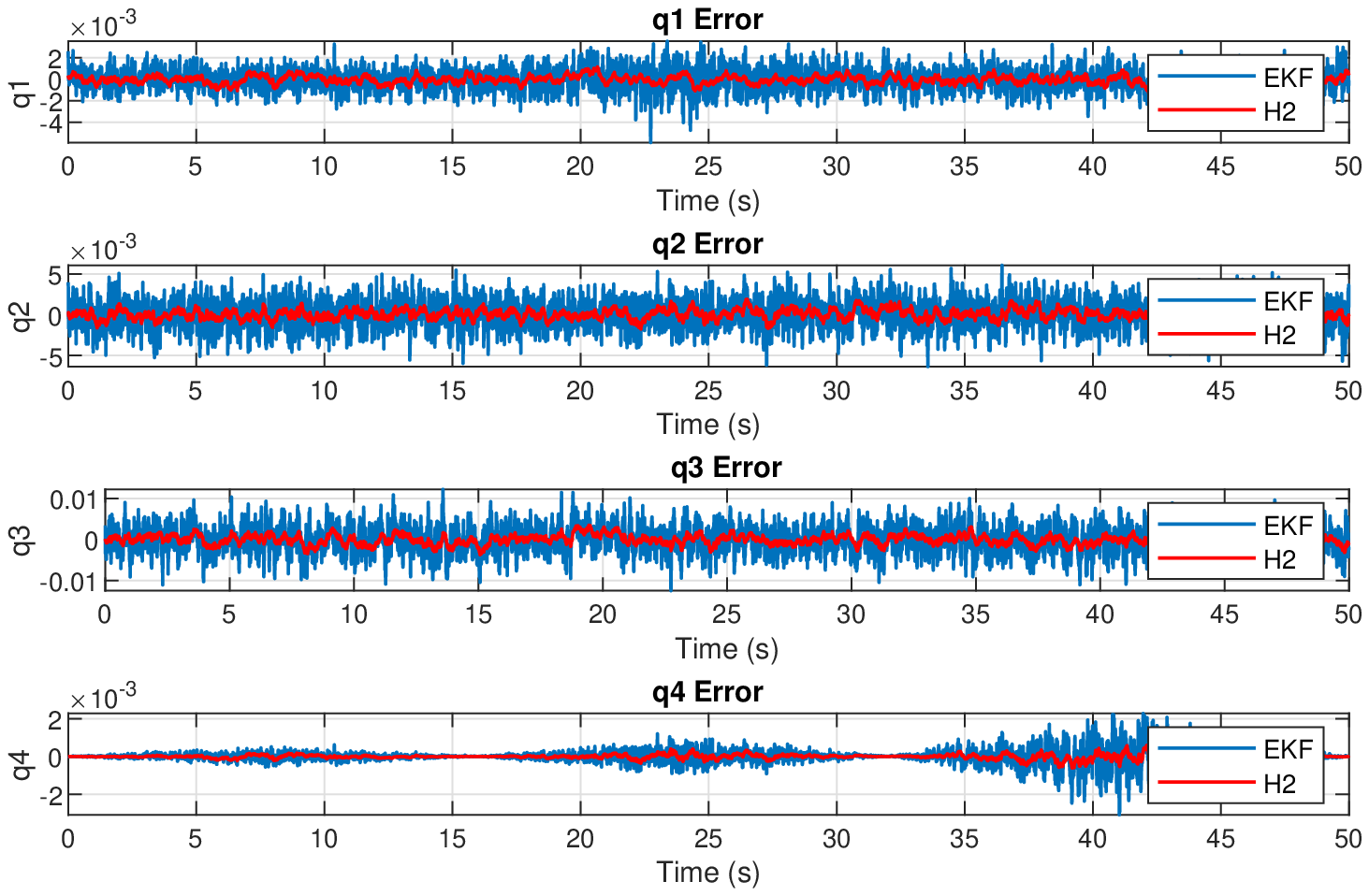}
    \caption{Comparison of the extended $\mathcal{H}_2$ filter and the EKF with quaternions  for Case I.}
    \label{Fig:Low_quat}
\end{figure}

\begin{figure}[h!]
\centering
    \includegraphics[width=0.8\textwidth]{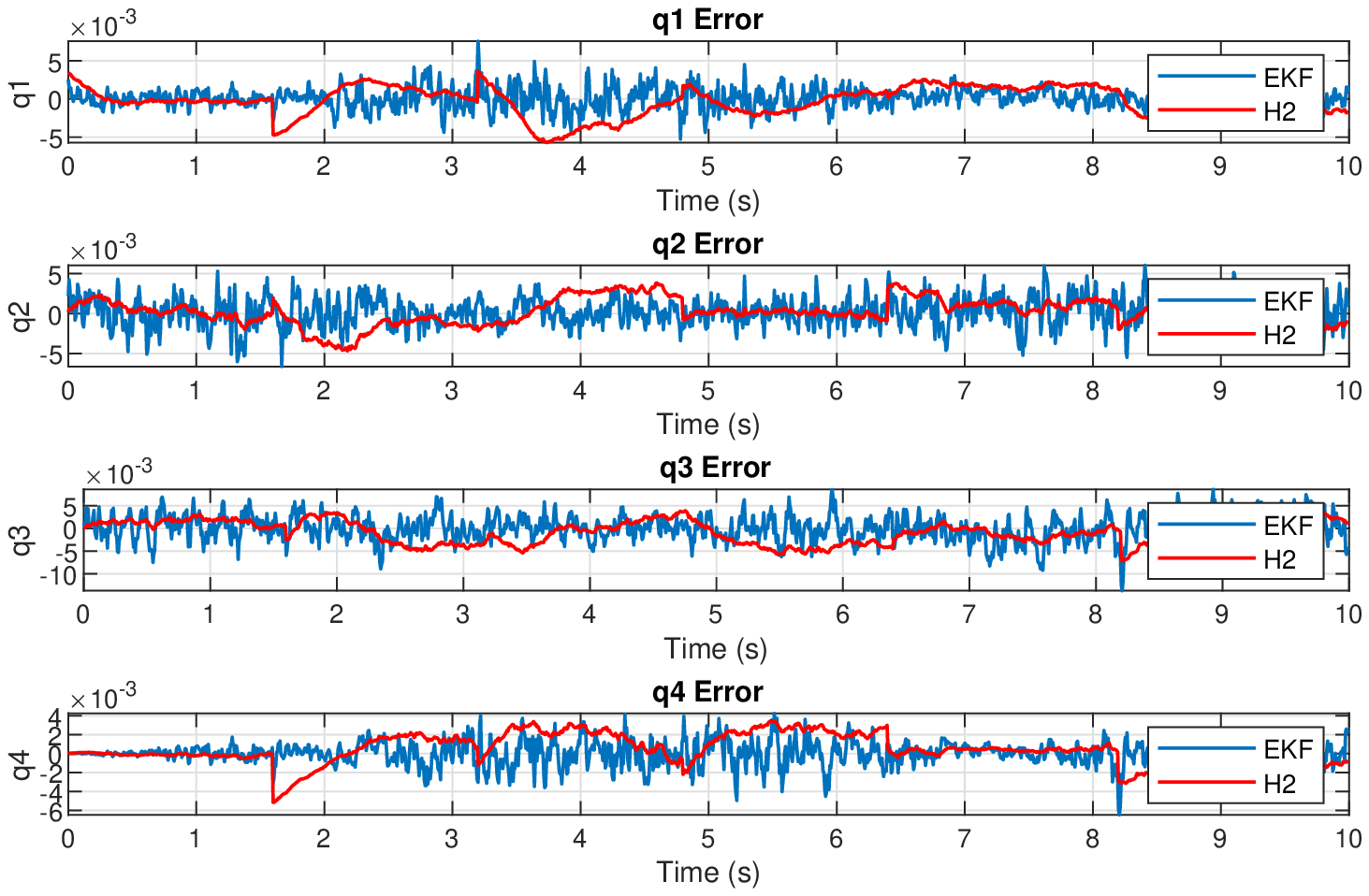}
    \caption{Comparison of the extended $\mathcal{H}_2$ filter and the EKF with quaternions for Case II.}
    \label{Fig:high_quat}
\end{figure}

\textbf{Case I and II:} The comparison of the two estimators for Case I and II is shown in Fig. \ref{Fig:Low_quat} and \ref{Fig:high_quat}, respectively. We can conclude that the performance of the extended $\mathcal{H}_2$ is slightly better than that of the EKF.

\begin{figure}[h!]
\centering
    \includegraphics[width=0.8\textwidth]{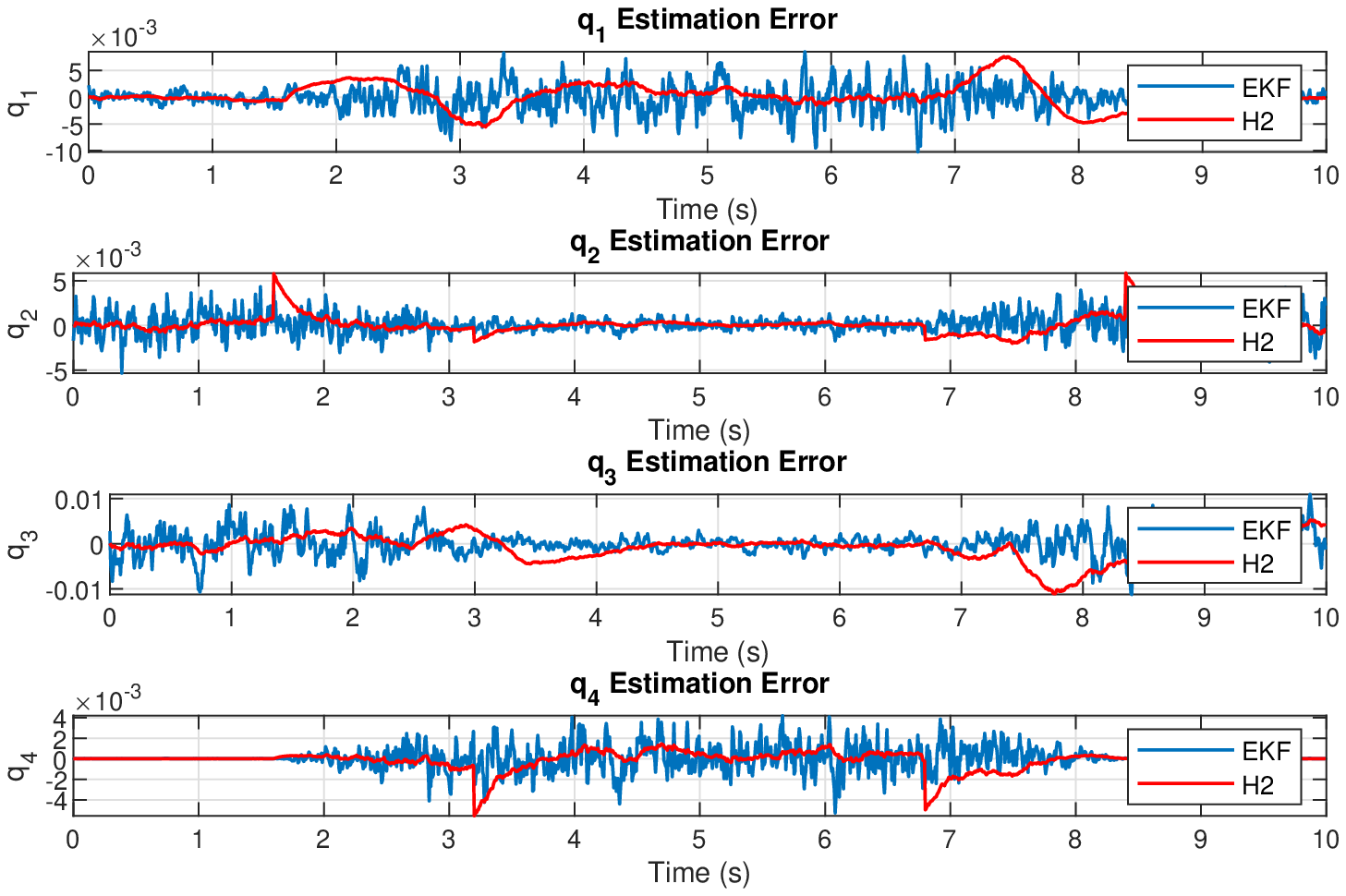}
    \caption{Comparison of the extended $\mathcal{H}_2$ filter and the EKF with quaternions for Case III.}
    \label{Fig:gimball_error}
\end{figure}

\textbf{Case III:} The comparison of the two estimators for Case III is shown in Fig. \ref{Fig:gimball_error}.

We observe that Extended $\mathcal{H}_2$ estimation is functional even when it encounters gimbal lock, a problem faced in an Euler angle-based implementation.
Moreover, the error of extended $\mathcal{H}_2$ estimator is comparable with that of EKF. The RMS error for both the filters are shown in Table \ref{Table:RMS_quat_gim}. From the plots and the data in the tables, we can conclude that the performance of extended $\mathcal{H}_2$ estimator is comparable with that of EKF.

\begin{table}[h]
\caption{RMS error for Case III.} \label{Table:RMS_quat_gim}
\begin{center}
\renewcommand{\arraystretch}{1.5}
\begin{tabular}{|c||c|c|}
\hline
Algorithm & Pitch angle(${}^\circ$) & Standard Deviation(${}^\circ$)\\
\hline \hline
Extended $\mathcal{H}_2$ & 0.1204 & 0.013\\
\hline
EKF & 0.1569 & 0.0143\\
\hline
\end{tabular}
\end{center}
\end{table}

\begin{figure}[h!]
\centering
    \includegraphics[width=0.8\textwidth]{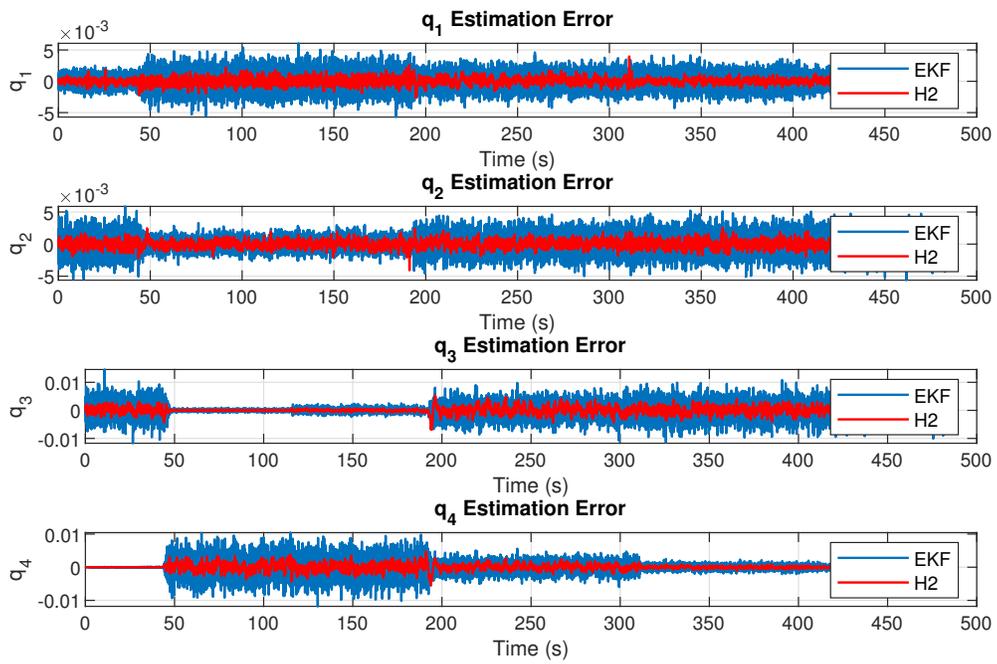}
    \caption{ Comparison of the extended $\mathcal{H}_2$ filter and the EKF for Case IV.}
\label{Fig:quat_compare_sim}
\end{figure}

\textbf{Case IV:} The comparison of the two estimators for Case IV is shown in Fig. \ref{Fig:quat_compare_sim}. We observe that the error of extended $\mathcal{H}_2$ estimator is comparable with that of EKF. The RMS error for both the filters are shown in Table \ref{Table:RMS_quat_sim}. From the plots and the data in the tables, we can conclude that the performance of extended $\mathcal{H}_2$ estimator is better than that of EKF for case IV as well.

\begin{table}[h]
\caption{RMS error for Case IV.} \label{Table:RMS_quat_sim}
\begin{center}
\renewcommand{\arraystretch}{1.5}
\begin{tabular}{|c||c|c|c|}
\hline
Algorithm & Roll angle(${}^\circ$) & Pitch angle(${}^\circ$) & Yaw angle(${}^\circ$)\\
\hline \hline
Extended $\mathcal{H}_2$ & 0.0410 & 0.0567 & 0.1274\\
\hline
EKF & 0.0956 & 0.1233 & 0.2761\\
\hline
\end{tabular}
\end{center}
\end{table}

\begin{table}[h!]
\caption{Computational Time Comparison on Quaternion Estimation} \label{Table:Com_time_Quat}
\begin{center}
\renewcommand{\arraystretch}{1.5}
\setlength{\tabcolsep}{10pt}
\begin{tabular}{|c||c|c|}
\hline
Algorithm & Mean Time ($ms$) & Standard Deviation ($ms$)\\
\hline \hline
Extended $\mathcal{H}_2$ &  0.9263 & 0.5217\\
\hline
EKF & 2.7 & 1.5 \\
\hline
\end{tabular}
\end{center}
\end{table}

\section{Conclusions}
This paper presents a new nonlinear estimation framework, based on $\mathcal{H}_2$ optimal state estimation, for attitude estimation in low power microprocessors.
This work showed that the performance of the proposed estimator is comparable, if not better, than that of the EKF algorithm which is typically used in the application space considered.  The primary advantage of the proposed framework is the $2\times$ computational efficiency, and the $3\times$ robustness with respect to computational uncertainty.
Both these factors make the proposed attitude estimation algorithm very attractive for small UAVs with low power microprocessors.

\section*{Funding Sources}
This project has been supported by NSF grants  \#1762825 and IUSE/PFE: RED:
REvolutionizing Diversity Of Engineering (REDO-E)
Award Number:1730693.




\bibliography{sample}

\begin{thebibliography}{27}
\newcommand{\enquote}[1]{``#1''}
\providecommand{\natexlab}[1]{#1}
\providecommand{\url}[1]{\texttt{#1}}
\providecommand{\urlprefix}{URL }
\expandafter\ifx\csname urlstyle\endcsname\relax
  \providecommand{\doi}[1]{\discretionary{}{}{}https://doi.org/#1}\else
  \providecommand{\doi}[1]{\discretionary{}{}{}\urlstyle{rm}\url{https://doi.org/#1}}\fi

\bibitem[{Eure et~al.(2013)Eure, Quach, Vazquez, Hogge, and
  Hill}]{eure2013application}
Eure, K.~W., Quach, C.~C., Vazquez, S.~L., Hogge, E.~F., and Hill, B.~L.,
  \enquote{An application of UAV attitude estimation using a low-cost inertial
  navigation system,} 2013.

\bibitem[{Gebre-Egziabher et~al.(2004)Gebre-Egziabher, Hayward, and
  Powell}]{gebre2004design}
Gebre-Egziabher, D., Hayward, R., and Powell, J., \enquote{Design Of
  Multi-sensor Attitude Determination Systems,} \emph{{IEEE} Transactions on
  Aerospace and Electronic Systems}, Vol.~40, No.~2, 2004, pp. 627--649.
\newblock \doi{10.1109/taes.2004.1310010},
  \urlprefix\url{https://doi.org/10.1109%2Ftaes.2004.1310010}.

\bibitem[{Kada et~al.(2016)Kada, Munawar, Shaikh, Hussaini, and
  Al-Saggaf}]{kada2016uav}
Kada, B., Munawar, K., Shaikh, M., Hussaini, M., and Al-Saggaf, U.,
  \enquote{{UAV} Attitude Estimation Using Nonlinear Filtering and Low-Cost
  Mems Sensors,} \emph{{IFAC}-{PapersOnLine}}, Vol.~49, No.~21, 2016, pp.
  521--528.
\newblock \doi{10.1016/j.ifacol.2016.10.655},
  \urlprefix\url{https://doi.org/10.1016%2Fj.ifacol.2016.10.655}.

\bibitem[{Weibel et~al.(2015)Weibel, Lawrence, and Palo}]{weibel2015small}
Weibel, D., Lawrence, D., and Palo, S., \enquote{Small Unmanned Aerial System
  Attitude Estimation for Flight in Wind,} \emph{Journal of Guidance, Control,
  and Dynamics}, Vol.~38, No.~7, 2015, pp. 1300--1305.
\newblock \doi{10.2514/1.g000888},
  \urlprefix\url{https://doi.org/10.2514%2F1.g000888}.

\bibitem[{Crassidis et~al.(2007)Crassidis, Markley, and
  Cheng}]{crassidis2007survey}
Crassidis, J.~L., Markley, F.~L., and Cheng, Y., \enquote{Survey of Nonlinear
  Attitude Estimation Methods,} \emph{Journal of Guidance, Control, and
  Dynamics}, Vol.~30, No.~1, 2007, pp. 12--28.
\newblock \doi{10.2514/1.22452},
  \urlprefix\url{https://doi.org/10.2514%2F1.22452}.

\bibitem[{Ludwig and Burnham(2018)}]{ludwig2018comparison}
Ludwig, S.~A., and Burnham, K.~D., \enquote{Comparison of Euler Estimate using
  Extended Kalman Filter, Madgwick and Mahony on Quadcopter Flight Data,}
  \emph{2018 International Conference on Unmanned Aircraft Systems ({ICUAS})},
  {IEEE}, 2018.
\newblock \doi{10.1109/icuas.2018.8453465},
  \urlprefix\url{https://doi.org/10.1109%2Ficuas.2018.8453465}.

\bibitem[{Teague(2016)}]{teague2016comparison}
Teague, H., \enquote{Comparison of Attitude Estimation Techniques for Low-cost
  Unmanned Aerial Vehicles,} 2016.

\bibitem[{Trawny and Roumeliotis(2005)}]{trawny2005indirect}
Trawny, N., and Roumeliotis, S.~I., \enquote{Indirect Kalman filter for 3D
  attitude estimation,} \emph{University of Minnesota, Dept. of Comp. Sci. \&
  Eng., Tech. Rep}, Vol.~2, 2005, p. 2005.

\bibitem[{Ko et~al.(2016)Ko, Jeong, and Bae}]{ko2016sine}
Ko, N., Jeong, S., and Bae, Y., \enquote{Sine Rotation Vector Method for
  Attitude Estimation of an Underwater Robot,} \emph{Sensors}, Vol.~16, No.~8,
  2016, p. 1213.
\newblock \doi{10.3390/s16081213},
  \urlprefix\url{https://doi.org/10.3390%2Fs16081213}.

\bibitem[{Jing et~al.(2017)Jing, Cui, He, Zhang, Ding, and
  Yang}]{jing2017attitude}
Jing, X., Cui, J., He, H., Zhang, B., Ding, D., and Yang, Y., \enquote{Attitude
  estimation for {UAV} using extended Kalman filter,} \emph{2017 29th Chinese
  Control And Decision Conference ({CCDC})}, {IEEE}, 2017.
\newblock \doi{10.1109/ccdc.2017.7979077},
  \urlprefix\url{https://doi.org/10.1109%2Fccdc.2017.7979077}.

\bibitem[{Madgwick(2010)}]{madgwick2010efficient}
Madgwick, S., \enquote{An efficient orientation filter for inertial and
  inertial/magnetic sensor arrays,} \emph{Report x-io and University of Bristol
  (UK)}, Vol.~25, 2010, pp. 113--118.

\bibitem[{Simon(2006)}]{simon2006optimal}
Simon, D., \emph{Optimal State Estimation}, John Wiley {\&} Sons, Inc., 2006.
\newblock \doi{10.1002/0470045345},
  \urlprefix\url{https://doi.org/10.1002%2F0470045345}.

\bibitem[{Kim et~al.(2020)Kim, Tadiparthi, and Bhattacharya}]{9147415}
Kim, S., Tadiparthi, V., and Bhattacharya, R., \enquote{Nonlinear Attitude
  Estimation for Small {UAVs} with Low Power Microprocessors,} \emph{2020
  American Control Conference ({ACC})}, {IEEE}, 2020.
\newblock \doi{10.23919/acc45564.2020.9147415},
  \urlprefix\url{https://doi.org/10.23919%2Facc45564.2020.9147415}.

\bibitem[{Choukroun et~al.(2011)Choukroun, Cooper, and
  Berman}]{choukroun2011spacecraft}
Choukroun, D., Cooper, L., and Berman, N., \enquote{Spacecraft Attitude
  Estimation and Gyro Calibration via Stochastic H
  {\hspace{0.167em}}$\infty${\hspace{0.167em}} Filtering,} \emph{Advances in
  Aerospace Guidance, Navigation and Control}, Springer Berlin Heidelberg,
  2011, pp. 397--415.
\newblock \doi{10.1007/978-3-642-19817-5_31},
  \urlprefix\url{https://doi.org/10.1007%2F978-3-642-19817-5_31}.

\bibitem[{Goldstein et~al.(2002)Goldstein, Poole, and
  Safko}]{goldstein2002classical}
Goldstein, H., Poole, C., and Safko, J., \enquote{Classical mechanics,} , 2002.

\bibitem[{Carrillo et~al.(2013)Carrillo, L{\'{o}}pez, Lozano, and
  P{\'{e}}gard}]{carrillo2013modeling}
Carrillo, L. R.~G., L{\'{o}}pez, A. E.~D., Lozano, R., and P{\'{e}}gard, C.,
  \enquote{Modeling the Quad-Rotor Mini-Rotorcraft,} \emph{Advances in
  Industrial Control}, Springer London, 2013, pp. 23--34.
\newblock \doi{10.1007/978-1-4471-4399-4_2},
  \urlprefix\url{https://doi.org/10.1007%2F978-1-4471-4399-4_2}.

\bibitem[{Allen and Chang(1993)}]{allen1993performance}
Allen, R., and Chang, D., \enquote{Performance testing of the systron donner
  quartz gyro,} \emph{Jpl Engineering Memorandum, EM}, 1993, pp. 343--1297.

\bibitem[{El-Sheimy et~al.(2008)El-Sheimy, Hou, and Niu}]{el2007analysis}
El-Sheimy, N., Hou, H., and Niu, X., \enquote{Analysis and Modeling of Inertial
  Sensors Using Allan Variance,} \emph{{IEEE} Transactions on Instrumentation
  and Measurement}, Vol.~57, No.~1, 2008, pp. 140--149.
\newblock \doi{10.1109/tim.2007.908635},
  \urlprefix\url{https://doi.org/10.1109%2Ftim.2007.908635}.

\bibitem[{Lam et~al.(2003)Lam, Stamatakos, Woodruff, and Ashton}]{lam2003gyro}
Lam, Q., Stamatakos, N., Woodruff, C., and Ashton, S., \enquote{Gyro Modeling
  and Estimation of Its Random Noise Sources,} \emph{{AIAA} Guidance,
  Navigation, and Control Conference and Exhibit}, American Institute of
  Aeronautics and Astronautics, 2003.
\newblock \doi{10.2514/6.2003-5562},
  \urlprefix\url{https://doi.org/10.2514%2F6.2003-5562}.

\bibitem[{Shuster(1993)}]{shuster1993survey}
Shuster, M.~D., \enquote{A survey of attitude representations,}
  \emph{Navigation}, Vol.~8, No.~9, 1993, pp. 439--517.

\bibitem[{Kuipers(1999)}]{kuipers1999quaternions}
Kuipers, J.~B., \emph{Quaternions and Rotation Sequences}, Princeton University
  Press, 1999.
\newblock \doi{10.1515/9780691211701},
  \urlprefix\url{https://doi.org/10.1515%2F9780691211701}.

\bibitem[{Duan and Yu(2013)}]{duan2013lmis}
Duan, G.-R., and Yu, H.-H., \emph{{LMIs} in Control Systems}, {CRC} Press,
  2013.
\newblock \doi{10.1201/b15060},
  \urlprefix\url{https://doi.org/10.1201%2Fb15060}.

\bibitem[{Apkarian et~al.(2001)Apkarian, Tuan, and
  Bernussou}]{apkarian2001continuous}
Apkarian, P., Tuan, H.~D., and Bernussou, J., \enquote{Continuous-time
  analysis, eigenstructure assignment, and H/sub 2/ synthesis with enhanced
  linear matrix inequalities ({LMI}) characterizations,} \emph{{IEEE}
  Transactions on Automatic Control}, Vol.~46, No.~12, 2001, pp. 1941--1946.
\newblock \doi{10.1109/9.975496},
  \urlprefix\url{https://doi.org/10.1109%2F9.975496}.

\bibitem[{Grant and Boyd(2014)}]{grant2009cvx}
Grant, M., and Boyd, S., \enquote{{CVX}: Matlab Software for Disciplined Convex
  Programming, version 2.1,} \url{http://cvxr.com/cvx}, mar 2014.

\bibitem[{Gu et~al.(2005)Gu, Petkov, and Konstantinov}]{gu2005robust}
Gu, D.-W., Petkov, P., and Konstantinov, M.~M., \emph{Robust control design
  with MATLAB{\textregistered}}, Springer Science \& Business Media, 2005.
\newblock \doi{10.1007/b135806},
  \urlprefix\url{https://doi.org/10.1007%2Fb135806}.

\bibitem[{InvenSense(2016)}]{MPU}
InvenSense, T., \enquote{MPU-9250 Product Specification Revision 1.1,} , 2016.
\newblock \urlprefix\url{http://www.invensense.com/wp-content/uploads
  /2015/02/PS-MPU-9250A-01-v1.1.pdf}, [Accessed: 13-Nov-2019].

\bibitem[{MATLAB(2018)}]{MAT}
MATLAB, \enquote{Sensor Fusion and Tracking Toolbox,} , 2018.
\newblock \urlprefix\url{https://www.mathworks.com/products/sensor-fu
  sion-and-tracking.html}, [Accessed: 18-Sep-2019].

\end{thebibliography}

\end{document}